\documentclass[fleqn,usenatbib]{mnras}

\usepackage{newtxtext,newtxmath}
\usepackage[T1]{fontenc}

\DeclareRobustCommand{\VAN}[3]{#2}
\let\VANthebibliography\thebibliography
\def\thebibliography{\DeclareRobustCommand{\VAN}[3]{##3}\VANthebibliography}

\usepackage{graphicx}	% Including figure files
\usepackage{amsmath}	% Advanced maths commands
\usepackage{pythonhighlight}
\usepackage{bm}
\usepackage{threeparttable}
\newcommand{\vect}[1]{\boldsymbol{\mathbf{#1}}}
\usepackage{multirow}
\usepackage{mathtools}
\usepackage{subfig}
\usepackage{xcolor}

\usepackage{xr}

\newcommand{\kms}{$\mathrm{km\,s^{-1}}$}

\newcommand{\kp}{$K_\mathrm{p}$}

\newcommand{\hdb}{HD\,189733\,b}
\newcommand{\pegb}{51\,Peg\,b}
\newcommand{\hd}{HD\,189733}
\newcommand{\peg}{51\,Peg}
\newcommand{\ninj}{$n_\mathrm{inj}$}

\title[GPs for High-Resolution Spectroscopy]{Applications of a Gaussian Process Framework for Modelling of High-Resolution Exoplanet Spectra}

\author[A. Meech]{
Annabella Meech$^{1}$\thanks{E-mail: annabella.meech@physics.ox.ac.uk},
Suzanne Aigrain$^{1}$,
Matteo Brogi$^{2}$,
Jayne L. Birkby$^{1}$
\\
$^{1}$Physics Department, University of Oxford, Denys Wilkinson Building, Oxford, OX1 3RH, UK\\
$^{2}$Physics Department, University of Warwick, Coventry, CV4 7AL, UK
}

\date{Accepted 2022 March 2. Received 2022 February 18; in original form 2021 October 12}

\pubyear{2021}

\begin{document}
\label{firstpage}
\pagerange{\pageref{firstpage}--\pageref{lastpage}}
\maketitle
\defcitealias{Brogi2013}{BR13}
\defcitealias{Birkby2013}{BI13}

\begin{abstract}
Observations of exoplanet atmospheres in high resolution have the potential to resolve individual planetary absorption lines, despite the issues associated with ground-based observations. The removal of contaminating stellar and telluric absorption features is one of the most sensitive steps required to reveal the planetary spectrum and, while many different detrending methods exist, it remains difficult to directly compare the performance and efficacy of these methods. Additionally, though the standard cross-correlation method enables robust detection of specific atmospheric species, it only probes for features that are expected \emph{a priori}. Here we present a novel methodology using Gaussian process (GP) regression to directly model the components of high-resolution spectra, which partially addresses these issues. We use two archival CRIRES/VLT data sets as test cases, observations of the hot Jupiters HD\,189733\,b and 51\,Pegasi\,b, recovering injected signals with average line contrast ratios of $\sim 4.37 \times 10^{-3}$ and $\sim 1.39 \times 10^{-3}$, and planet radial velocities $\Delta K_\mathrm{p} =1.45 \pm 1.53$\,\kms{} and $\Delta K_\mathrm{p}=0.12\pm0.12$\,\kms{} from the injection velocities respectively. In addition, we demonstrate an application of the GP method to assess the impact of the detrending process on the planetary spectrum, by implementing injection-recovery tests. We show that standard detrending methods used in the literature negatively affect the amplitudes of absorption features in particular, which has the potential to render retrieval analyses inaccurate. Finally, we discuss possible limiting factors for the non-detections using this method, likely to be remedied by higher signal-to-noise data.
\end{abstract}

\begin{keywords}
atmospheric effects -- infrared: planetary systems -- methods: data analysis -- planets and satellites: atmospheres -- techniques: spectroscopic
\end{keywords}

%%%%%%%%%%%%%%%%%%%%%%%%%%%%%%%%%%%%%%%%%%%%%%%%%%

%%%%%%%%%%%%%%%%% BODY OF PAPER %%%%%%%%%%%%%%%%%%

\section{Introduction}
\label{sec:1 intro}
High-resolution spectroscopy (HRS) is one of the current leading techniques used to probe exoplanet atmospheres and, since its conception \citep{Snellen2010}, has enabled inference of the presence of atomic and molecular species, atmospheric dynamics and temperature structure \citep{Brogi2012, Rodler2012, Birkby2017, Nugroho2017, Flowers2019, Ehrenreich2020, Pino2020, Beltz2021,Wardenier2021}. From the Earth we observe the composite spectrum
\begin{equation}
    F_{\mathrm{obs}} = (F_* + F_\mathrm{p}) \times T \times A,
\label{eqn:composite spectrum}
\end{equation}
\noindent where $F_*$, $F_\mathrm{p}$ and $T$ are the stellar, planet, and telluric spectra, respectively, altered by the instrumental transmission $A$, with each component a function of time and wavelength. $F_*$ and $F_\mathrm{p}$ shift according to the stellar reflex motion and planet orbital motion, respectively, the latter typically orders of magnitude greater. The principal hurdle is distinguishing the relatively weak planet signal from the multitude of nuisance signals, the dominant contributions being the stellar and variable telluric absorption. Though occasionally individual planet atmospheric absorption lines can be distinguished \citep{Schwarz2016,Brogi2012}, they typically have a signal-to-noise ratio (SNR) smaller than 1, even when combining multiple nights of data. The key insight is that the planet's rest frame moves significantly with respect to the other components given in equation~\ref{eqn:composite spectrum}. Therefore, if observed at different orbital phases, we can isolate the exoplanet's transmission or emission spectrum by its Doppler shift.

One of the prevailing challenges to maximizing HRS capabilities is the effective removal of telluric contamination, while preserving the true astronomical signal. Though telluric signals are considered stationary in wavelength, the rapid fluctuations of the Earth's atmospheric conditions cause variation in depth and shape over short time-scales. Observing bands in the near-infrared (NIR) offer a more favourable planet-to-star contrast ratio; however, there are regions of strong telluric absorption at these wavelengths due to dominant species CH$_4$, CO$_2$, and H$_2$O. Many different telluric detrending techniques have been successfully implemented for HRS. Theoretical atmospheric transmission modelling, with codes such as \textsc{telfit} and \textsc{molecfit} \citep{Gullikson2014, Smette2015}, combines a radiative transfer treatment of the Earth's atmosphere with observatory metadata to build a synthetic telluric model. Though very effective in the optical \citep{Casasayas-Barris2019, Hoeijmakers2020, Bourrier2020}, such modelling fails to capture variations in the heavy absorption seen at longer wavelengths, thus empirical methods have been favoured in NIR studies. Some have opted for principal component analysis (PCA) to remove contaminating features: \citet{Dekok2013} were one of the first to apply this decomposition of the spectral matrix in order to identify static telluric features. Later \citet{Piskorz2016,Piskorz2017,Piskorz2018} applied a similar technique to NIRSPEC/Keck observations, guiding the PCA with a baseline telluric model. The recent study of HD\,209508\,b then further developed the PCA technique to detrend GIANO-B/TNG spectra \citep{Giacobbe2021}. SYSREM, a specific, data point uncertainty-weighted PCA-based algorithm \citep{Tamuz2005,Mazeh2007}, has been widely used to correct for common mode systematic effects in high-resolution spectral data, and has proved to be a powerful tool, particularly for removal of strong contamination \citep{Birkby2013,Nugroho2017,Gibson2020,Kesseli2020,Merritt2020}. For each pass, the SYSREM algorithm produces a low-rank approximation of the spectral matrix, a product of two column vectors, identifying common modes in each wavelength channel that can then be subtracted (see \citealt{Tamuz2005} for further detail). The low-order trends are often correlated with physical conditions such as airmass and seeing \citep{Birkby2017}. SYSREM requires fine-tuning to carefully remove as much of the telluric contamination as possible without infringing on excessive use, detrimental to the planet spectrum. Another popular technique uses linear regression to model the variation of flux with airmass, combined with additional sampling of particular wavelengths known to accommodate strong telluric lines \citep{Brogi2012,Brogi2013,Brogi2014,Webb2020}. In this method the fluxes in each wavelength channel (pixel) are divided through by a least-squares fitted linear trend with airmass. This first pass captures the low-order trends of the telluric variation. Then, the evolution of residuals in each individual pixel is modelled linearly with the temporal variation of the (summed) residuals at locations of known strong telluric lines. Finally, a high-pass filter is applied, involving outlier rejection and removal of any remaining trends in the spectral direction. While SYSREM searches for any common modes in wavelength, including any quasi-stationary stellar lines, the assumption of airmass variation in the latter method makes it unsuitable when strong stellar features are present.

Given the plethora of techniques that can be used to remove telluric contamination, it can be difficult to choose the most suitable approach for a particular data set, considering the atmospheric conditions on the night of observation, and spectral range. Some studies have used the cross-correlation detection significance to compare different detrending algorithms \citep{Cabot2019,Langeveld2021}. Moreover, for atmospheric characterization we are concerned with the impact on the planet spectrum. It is widely accepted that most detrending algorithms alter the planet signal in some way; hence it is important to subject the cross-correlation models to the same processing for accurate retrieval work \citep{Brogi2019,Gibson2020}. However, since the planet spectrum goes unseen with these standard methods, it is not straightforward to pinpoint the modifications.

High-resolution cross-correlation spectroscopy (HRCCS) enables detection of species by way of a line-matching exercise: the residual, corrected fluxes are cross-correlated with a template planetary spectrum, derived by assuming a radiative transfer treatment, temperature profile and abundance of chemical species in the planet's atmosphere. The planet signal is typically much weaker than the noise of the detrended fluxes, and even the photon noise level. Since the SNR of the planet is proportional to the number of detected lines \citep{Birkby2018}, HRCCS combines the signal from each line: the cross-correlation functions (CCFs) resulting from each spectrum are aligned into the assumed planet rest frame and summed. This is repeated for a range of systemic and planet radial velocities. Then, there are various metrics used to compute the detection significance from the cross-correlation velocity maps, notably comparing the signal of the cross-correlation peak to the standard deviation of `noise' outside the peak, and separately the Welch \emph{t}-test, whereby the null hypothesis, that the CCFs within the planet trail are drawn from the same parent distribution as those outside the trail is rejected at a certain significance \citep{Brogi2013, Dekok2013, Birkby2013, Birkby2017}. Both methods have their known flaws when it comes to deriving reliable detection significances \citep{Cabot2019} and translating to estimates of atmospheric properties. There has been work to map the cross-correlation values to a log-likelihood, in order to enable integration into a Bayesian analysis retrieval framework \citep{Brogi2019,Gibson2020}, akin to low-resolution spectroscopy retrievals \citep{Madhusudhan2009}. These approaches then allow more statistically robust model testing by exploration of the full parameter space, and have higher constraining power. That said, they assume that the data are independent, and the likelihood of \citet{Gibson2020} assumes that all data point uncertainties are uniformly scaled. This may be problematic in some cases, for example when dealing with both regions of continuum and cores of telluric lines.

In this work, we present a Gaussian process (GP) regression methodology to directly model the telluric and planet signals sequentially, analyse its potential, and compare its efficacy to the results of current popular approaches. In section~\ref{sec:2 gp modelling}, we present the GP framework used to forward model the composite spectrum. We investigate the performance of sequential GP modelling by application to archival NIR observations of the dayside of two well-studied hot Jupiters, HD\,189733\,b and 51\,Peg\,b; the data reduction and results of these tests are described in section~\ref{sec:3 test data}. We then discuss an application of the method, namely the comparison of different telluric corrections in section~\ref{sec:4 impact of telluric methods}, and our conclusions are outlined in section~\ref{sec:5 conclusions}.
\section{Proposed GP Modelling of High-Resolution Spectra}
\label{sec:2 gp modelling}
There have been many successful examples of applying GPs to time-series exoplanet data, modelling stellar variability for exoplanet radial velocity extraction, stellar-rotation periods, photometry, and systematics for (e.g.) low-resolution transmission spectroscopy \citep{Rajpaul2015,Angus2018,Aigrain2016,Gibson2012}. \citet{Czekala2017} first presented the concept of employing GPs to disentangle different components in high-resolution spectra, specifically binary stellar spectra, with the package \textsc{psoap}. Their method was limited by computational expense, only reasonably allowing evaluation of the GP on narrow bandpasses and requiring access to computer clusters. We propose that GP regression offers an attractive route for direct inference of the components of exoplanet high-resolution spectra in a principled, probabilistic manner. Here we attempt to create a reasonably fast GP framework for this purpose.

\subsection{Gaussian process regression}
\label{sec:2 gpr}
We give a brief introduction of GPs and regression for the context of the methods presented in this paper - for a fuller description please see \citet{Rasmussen}. A Gaussian process is a special case of stochastic process based on a multivariate Gaussian distribution (extended to an infinite number of variables), and is useful for both regression and classification. In the context of regression, GPs allow us to infer probability distributions over non-parametric functions, affording highly flexible models, and are tractable since we are only concerned about evaluation at a finite number of outputs. We define the properties of the functions modelled by a GP via a mean function $m(\mathbf{x})$ and a covariance function $k(x_i,x_j)$, incorporating any prior knowledge of the form of the functions. We use the chosen kernel, $k(x_i,x_j)$, to compute covariance between any two inputs, thereby constructing the covariance matrix $\mathbf{K}$ of the GP. For an input vector $\mathbf{x}$, we assume our observations $\mathbf{y}$ are drawn from a multivariate Gaussian distribution: 
\begin{equation}
    p(\mathbf{y}) = \mathcal{N}(\mathbf{m},\mathbf{K}).
\label{eqn: gp prior}
\end{equation}

\noindent
It follows that the GP log-likelihood for $N$ observations $\mathbf{y}$ is given by
\begin{equation}
    \log p(\mathbf{y}|\vect{x,\phi,\theta}) = -\frac{1}{2}\mathbf{(y-m)}^\mathrm{T}\mathbf{K}^{-1}\mathbf{(y-m)}-\frac{1}{2}\mathrm{log}|\mathbf{K}|-\frac{N}{2}\mathrm{log}2\pi
\label{eqn: gp loglike}
\end{equation}
where $\vect{\phi}$ and $\vect{\theta}$ are the parameters of the mean and covariance functions, thus hyperparameters of the GP. By evaluation of this likelihood, we first sample the posterior distributions of the hyperparameters 
\begin{equation}
    p(\vect{\phi,\theta}|\mathbf{y}) \propto p(\mathbf{y}|\vect{\phi,\theta})p(\vect{\phi,\theta}),
\label{eqn:posterior dist}
\end{equation}
\noindent where $p(\vect{\phi,\theta})$ is the prior on the hyperparameters. Having trained the GP, we are interested in prediction at test locations $\mathbf{x_*}$. Since the training and test sets are jointly Gaussian distributed, the predictive posterior distribution can be derived as
\begin{equation}
    \mathrm{p}(\mathbf{y_*}|\mathbf{y},k) = \mathcal{N}(\mathbf{K_*^\mathrm{T}}\mathbf{K}^{-1}\mathbf{y},\mathbf{K_{**}}-\mathbf{K_*^\mathrm{T}}\mathbf{K}^{-1}\mathbf{K_*})
\label{eqn: predictive dist}
\end{equation}
\noindent for $\mathbf{m = 0}$, where $\mathbf{K_{**}} = k(\mathbf{x_*,x_*})$, the covariance function evaluated for pairs of test points and $\mathbf{K_{*}} = k(\mathbf{x,x_*})$ that for pairs of training and test points.

%%%%%%%%%%%%%%%%%%%%%%%%%%%%%%%%%%%%%%%%%%%%%%%%%%%%%%%%%%%%%%%%%%%%%%%%%%%%%%%%%%%%%%%%%%%%%%%%%%%%%%%%
%%%%%%%%%%%%%%%%%%%%%%%%%%%%%%%%%%%%%%%%%%%%%%%%%%%%%%%%%%%%%%%%%%%%%%%%%%%%%%%%%%%%%%%%%%%%%%%%%%%%%%%%

\subsection{Application of GP regression to high-resolution spectroscopy}
\label{sec:2 full method hrs}
Although some underlying physics concerning possible absorbing species, temperature structure and cloud formation is established, it is difficult to construct a suitable (physically motivated) model parametrization for high-resolution exoplanet spectra. By modelling with a GP we acknowledge that there exists some correlation between flux uncertainties in wavelength, albeit locally, which can be specified via a parametrized covariance function. We assume little about the functional form of the spectra \emph{a priori}, modelling each component spectrum of equation~(\ref{eqn:composite spectrum}) as 
\begin{equation}
    f \sim \mathcal{GP}(\vect{0},\vect{K(\lambda,\theta)})
\label{eqn:high res gp form}
\end{equation}
\noindent where $f$ are the spectral fluxes (e.g. $F_\mathrm{p}$) and $\vect{\lambda} = (\lambda_1,...,\lambda_N)^\mathrm{T}$ is the distinct vector of wavelengths in the rest frame of the particular component. In theory, we could use a multidimensional Gaussian process to forward model the composite spectrum $F_\mathrm{obs}$, as given in equation~(\ref{eqn:composite spectrum}).

Though a Gaussian process is not a particularly natural choice of model for a spectrum, such an approach offers a number of anticipated benefits. First, whereas HRCCS only allows detection of the atmospheric chemical species included in the cross-correlation template, the proposed GP method affords an estimate of the functional form of the planetary spectrum, potentially facilitating detections of unsought species (and therefore unforeseen physics). In principle, this planetary spectrum could be directly fed into a retrieval framework in a similar way to low-resolution retrievals. That said, bypassing some of the available information, such as the location of planetary absorption features, is likely to cause a reduction in sensitivity compared to HRCCS. Further, using this methodology we are able to robustly propagate uncertainty estimates in a principled manner. This was a prevalent shortfall with standard HRCCS techniques, though we note that the cross-correlation to likelihood method now allows it, albeit in cross-correlation space \citep{Brogi2019, Gibson2020}. Analyses of high-resolution spectra also require continuum normalization for self-calibration. This results in the loss of broad-band variations and the planetary continuum, thus all measurements of absorption and emission features are relative. Though (e.g.) \citet{Brogi2019} and \mbox{\citet{Line2021}} show that there still remains enough information in the spectra to retrieve absolute abundances and temperature-pressure profiles, retaining the absolute planetary continuum would further help to constrain these atmospheric properties. Lastly, using GPs in this context allows modelling of correlations between data points; HRCCS methods assume the data are independent.

%%%%%%%%%%%%%%%%%%%%%%%%%%%%%%%%%%%%%%%%%%%%%%%%%%%%%%%%%%%%%%%%%%%%%%%%%%%%%%%%%%%%%%%%%%%%%%%%%%%%%%%%
\subsection{Sequential GP modelling}
\label{sec:2 sequential method}
The number of operations required to compute the inverse and log-determinant of $\mathbf{K}$, needed for $\log \mathcal{L}$ in equation~(\ref{eqn: gp loglike}), scales as $\mathcal{O}(N^3)$. This is a limiting factor considering the typically large $N$ for high-resolution observations. To mitigate this challenge, we use the \textsc{python} GP implementation \textsc{celerite} for our GP models, which makes use of specific, stationary forms of the covariance function for faster, $\mathcal{O}(N)$ evaluation \citep{DFM2017}. A stationary kernel is one that depends only on relative positions in input space rather than absolute positions. The form of the \textsc{celerite} kernel limits it to 1D inputs\footnote{\citealt{Gordon2020} extend \textsc{celerite} to 2D inputs, but only for very specific conditions that do not apply to the present problem.}. We therefore present a method to construct a model for and remove each component separately in this work, namely the Earth's transmission spectrum, $T(\vect{\lambda}_1,t)$, first and then the planetary absorption (or emission) spectrum, $F_\mathrm{p}(\vect{\lambda}_2)$ (see equation~\ref{eqn:composite spectrum}). Not only does this sequential method overcome the computational limitation, but also introduces modularity, allowing us to test each successive step in turn. We note that this method does not yet preserve the continuum, since we normalize the spectra prior to modelling. We assume that:
\begin{enumerate}
    \item the spectral components are each a realization of an independent GP;
    \item the spectra are perfectly normalized and wavelength calibrated, and the instrument line spread function (LSF) is stable, so we are not concerned with instrumental effects;
    \item the tellurics vary only with airmass;
    \item the planet spectrum, $F_\mathrm{p}$ is constant in time;
    \item the stellar spectrum has been adequately removed prior to implementation of the GP framework.
\end{enumerate}
In practice, these assumptions are unlikely to be strictly correct. We now proceed to examine each one in turn and discuss the likelihood and impact of it being violated.
\begin{enumerate}
    \item The assumption that each component can be treated as a GP has been addressed above already. Their independence is not in question, provided we are considering thermal emission. For reflected light, the planet's spectrum will depend on the star's spectrum, but this is not the case for the data sets considered in this work.
    \item If the wavelength calibration is significantly erroneous, this could lead to reduced or missed feature recovery. That said, this is a standard assumption in HRCCS. An unstable LSF would cause a variation in resolution of the observed planet spectrum with time, so should be assessed for each data set. The measured variation in resolving power for these data do not significantly affect the recovered GP spectrum.
    \item While airmass is the dominant parameter controlling the time-dependence of telluric absorption, other factors are also important, such as variable precipitable water vapour. Furthermore, the dependence on airmass is no longer linear in the strong absorption regime. All in all, this is probably the most problematic assumption in our framework, and the reason why we also investigate other telluric correction methods in Section~\ref{sec:3 other detrending}.
    \item The planet's spectrum may change in time, particularly as the dayside rotates into view. First of all, the overall planet-to-star contrast should change, as the dayside of the planet rotates into or out of view. This would not be a problem when using the cross-correlation method to identify the planet's spectral features, but might need to be taken into account when forward-modelling the star and planet spectra simultaneously. Modelling them sequentially alleviates this problem, as we essentially lose the continuum information. Furthermore, as the day- and nightside can have very different temperatures, this can result in significant changes in atmospheric structure and composition, which would affect the features in the spectrum. We expect negligible variation in these data, considering the duration of observations and the relatively long rotational periods.
    \item This assumption was addressed in detail by \citet{Chiavassa2019}. For the purposes of this work, while residual stellar features may impede our ability to recover the planetary spectrum to some extent, the effect of residual tellurics is much more significant. 
\end{enumerate}

Appropriate choice of kernel is important for any GP model; \citet{Rasmussen} offer an extensive guide on model selection in ch. 5. For astrophysical applications it is preferable to consider the physical processes producing the signal when choosing a suitable kernel. Atomic and molecular absorption is a somewhat predictable process but variation in abundance along the line of sight, weather conditions and presence of clouds and hazes can vary the occurrence, breadth and strength of absorption. In our case, the spectral features are altogether modelled via the covariance function (since $\vect{m}=\vect{0}$). The Matern class of covariance functions, $k_\nu(x_i,x_j)$ where $\nu$ defines the degree of differentiability of the output functions, produces somewhat rough behaviour compared to other widely used kernels such as the squared exponential kernel. \citet{Rajpaul2020} assessed kernels for modelling of stellar spectra, finding the Matern-5/2 kernel to offer reasonable flexibility and smoothness. We choose to use the slightly rougher Matern-3/2 kernel because it (marginally) outperformed other kernels, available in the current \textsc{celerite} library, in a series of injection-recovery tests. It takes the form
\begin{equation}
    k_{3/2}(\lambda_i,\lambda_j) = \sigma^2 (1+ \frac{\sqrt{3}}{\rho}|\lambda_i-\lambda_j|)\exp{(-\frac{\sqrt{3}}{\rho}|\lambda_i-\lambda_j|)},
\label{eqn: matern32 kernel}
\end{equation}
where $\vect{\theta}=(\sigma,\rho)$ are hyperparameters corresponding to an amplitude and length scale. For this kernel the covariance between data points separated by $\rho$ in input space falls by 50\%. While we use the same kernel for both components (subsequently discussed) in this work, this is not obligatory for either the sequential method presented here or a hypothetical, future implementation where multiple components are modelled simultaneously.

%%%%%%%%%%%%%%%%%%%%%%%%%%%%%%%%%%%%%%%%%%%%%%%%%%%%%%%%%%%%%%%%%%%%%%%%%%%%%%%%%%%%%%%%%%%%%%%%%%%%%%%%
%%%%%%%%%%%%%%%%%%%%%%%%%%%%%%%%%%%%%%%%%%%%%%%%%%%%%%%%%%%%%%%%%%%%%%%%%%%%%%%%%%%%%%%%%%%%%%%%%%%%%%%%
\subsubsection{The telluric component}
\label{sec:2 telluric component}
High-resolution spectrographs are typically ground-based instruments, hence observations are subject to contamination by the Earth's transmission spectrum. The dominant absorbers are H$_2$O, OH and O$_2$, though other species such as CH$_4$ and CO$_2$ contribute significantly in the NIR. It is vital to remove imprinted features from the spectra accurately to distinguish the true planet signals. The telluric spectrum is variable over a period of HRS observations due to weather conditions, variable water vapour abundance and airmass. Assuming a plane-parallel atmosphere, telluric transmission follows
\begin{equation}
    T(\lambda,t) = \frac{I}{I_0} = \mathrm{exp}({-\tau(\lambda)a(t)}),
\label{eqn:tell transmission}
\end{equation}
\noindent where $I(\lambda)$ is the intensity, $I_0(\lambda)$ that above the atmosphere, $a(t)$ the airmass at time $t$ and $\tau$ the optical depth at zenith ($a=1$) \citep{Noll2012}. We therefore construct a telluric model 
\begin{equation}
    T(\lambda, t) = T_\mathrm{ref}(\lambda)^{a(t)},
\label{eqn:tell model}
\end{equation}
\noindent where $T_\mathrm{ref}(\lambda)$ is the Earth's transmission at zenith. A widely used technique in the field, which assumes the same treatment of tellurics, builds $T_\mathrm{ref}$ via linear regression to derive $\tau$ \citep{Wyttenbach2015, Astudillo-Defru2013, Vidal-Madjar2010}. Here, we model it with a GP. Given exponentiation is not an affine transform, we model the tellurics in log space. While we recognize that the propagated uncertainties (in log space) are no longer Gaussian, this does not have a significant impact on the result, likely due to the flexibility of GP models. We evaluate $\mathrm{log}\mathcal{L}$ (equation~\ref{eqn: gp loglike}) on the time-average of observations $\vect{y}~=~\mathrm{log}(F_{\mathrm{obs}}) / a(t)$, assuming all exposures share a common wavelength solution in the Earth's rest frame, and use the maximum likelihood estimate (MLE) to set the values of $\vect{\theta} = \{\sigma,\rho\}$. We treat each detector separately since the level of telluric absorption is expected to vary between them. For the newly defined $\vect{K}$, we compute the predictive distribution (equation~\ref{eqn: predictive dist}), for $\vect{x_*}$ defined at each pixel on the detector. The predictive mean $\vect{\mu_*}$ is then used to construct $T_\mathrm{ref}= \mathrm{exp}({\vect{\mu_*}})$.

%%%%%%%%%%%%%%%%%%%%%%%%%%%%%%%%%%%%%%%%%%%%%%%%%%%%%%%%%%%%%%%%%%%%%%%%%%%%%%%%%%%%%%%%%%%%%%%%%%%%%%%%
%%%%%%%%%%%%%%%%%%%%%%%%%%%%%%%%%%%%%%%%%%%%%%%%%%%%%%%%%%%%%%%%%%%%%%%%%%%%%%%%%%%%%%%%%%%%%%%%%%%%%%%%
\begin{table*}
\begin{center}
\begin{threeparttable}
\caption{Summary of test data sets analysed in Section~\ref{sec:3 test data}, where $t_\mathrm{exposure}$ is the exposure time, $M_\mathrm{spectra}$ is the number of separate exposures and $N_\mathrm{SYSREM}$ gives the optimal number of SYSREM iterations found in Section~\ref{sec:4 sysrem tuning} for each of the four detectors.}
\begin{tabular}{|l|c|c|c|c|c|} 
\hline
Object & Date & Phase coverage & $t_\mathrm{exposure}$ (s) & $M_\mathrm{spectra}$ & $N_\mathrm{SYSREM}$\\
\hline
HD\,189733\,b & 2011-08-01 & 0.383-0.475 & 150 & 48 & [4, 5, 4, 4] \\ 
51\,Peg\,b & 2010-10-16 & 0.36-0.42 & 42 & 166 & [3, 3, 4, 2] \\ 
 & 2010-10-17 & 0.60-0.66 & 42 & 148 & [2, 1, 5, 1] \\
 & 2010-10-25 & 0.49-0.54 & 42 & 138 & [2, 5, 3, 3] \\
\hline
\label{table:observations}
\end{tabular}
\end{threeparttable}
\end{center}
\end{table*}

\subsubsection{Modelling the planet signal}
\label{sec:2 planet fit}
Once the spectra have been corrected for telluric absorption we attempt to find the planet signal within the noise of the residuals.
We compute the equivalent wavelengths of each spectrum in the rest frame of the planet, having considered corrections for the Solar system barycentric velocity ($v_\mathrm{bary}$), velocity of the observed system ($v_\mathrm{sys}$) and the planet's radial velocity. For clarity, the resulting Doppler shift is given by the total planet velocity
\begin{equation}
    v_\mathrm{p} = v_\mathrm{bary}+v_\mathrm{sys}+ K_\mathrm{p}\mathrm{sin}(2\pi\varphi),
\label{eqn: vp}
\end{equation}
where $\varphi$ are the planet orbital phases and \kp{} is the semi-amplitude of the planet radial velocity signature. Rather than analyse each spectrum individually, we combine the shifted spectra from all spectral orders to produce a composite `spectrum' in the rest frame of the planet, by concatenating with the proper wavelengths in the planet rest frame. We model the planet spectrum with a second, independent GP. Though we are able to fit for \kp{}, this method is not sensitive to the constant $v_\mathrm{sys}$ since we only work with relative wavelength shifts, with no knowledge of the position of features. We place flat priors on all three free parameters, \kp{}$\sim \mathcal{U}(100,200)$\,\kms, $\sigma\sim \mathcal{U}(0,\infty)$, and $\rho\sim \mathcal{U}(\mathrm{d}\lambda, \infty)$ where $\mathrm{d}\lambda$ is the spectrograph resolution, and compute the GP log-likelihood (see equation~\ref{eqn: gp loglike}). Exploration of the full joint posterior distribution [e.g. using a Markov Chain Monte Carlo (MCMC) algorithm such as \textsc{emcee} \citep{DFM2013}] enables estimates of the three parameters with uncertainties. We use these estimates to construct the covariance matrix and finally condition the GP, obtaining the predictive mean planet spectrum. Since we do not expect significant correlation between test points located far apart in wavelength space, we condition the GP on the subset of data within 10$\rho$ of each test position.

\section{Tests on NIR archival data}
\label{sec:3 test data}

\subsection{Data format and prior reduction}
\label{sec:3 data reduction}
To test our GP framework, we reanalyse archival observations of the widely studied HD\,189733\,b and 51\,Peg\,b, taken on 2011 August 1 and over three nights, 2010 October 16th, 17th, and 25th, respectively. We attempt to recover the previously published, significant detections obtained by the standard methods introduced section~\ref{sec:1 intro}. Table~\ref{table:observations} summarizes the details of the dayside observations; we refer the reader to the original analyses of these data in \citet[][hereafter BI13]{Birkby2013} and \citet[][hereafter BR13]{Brogi2013}. In both cases observations were taken with the CRyogenic Infra-Red Echelle Spectrograph (CRIRES; \citealt{Kaeufl2004}) on the Very Large Telescope, as part of the large ESO Program 186.C-0289. CRIRES imaged the spectra via four Aladdin $\rm{II}$ detectors, each $1024\times512$ pixels with gaps between each detector, with wavelength coverages $3.1805-3.2659\,\mu$m and $2.287-2.345\,\mu$m, and a resolution of $R = \lambda/\Delta \lambda \simeq 100,000$. Extraction of the spectra and basic data reduction was completed via the CRIRES pipeline, v2.2.1 and v1.11.0 for the HD\,189733\,b and 51\,Peg\,b observations, respectively.

In both cases the original authors grouped the spectra into matrices $M$ spectra by $N$ pixels, one for each of the four detectors on CRIRES. In this work we begin the data processing and implementation of the GP framework with the pre-processed spectra, which were previously flat-fielded, bad pixel corrected and background subtracted. We note that the spectra had also been aligned to a common wavelength grid in the Earth's rest frame; \citetalias{Birkby2013} used the highest SNR spectrum as a reference with which to cross-correlate and \citetalias{Brogi2013} compared the positions of the centres of telluric lines with the average spectrum.

%%%%%%%%%%%%%%%%%%%%%%%%%%%%%%%%%%%%%%%%%%%%%%%%%%%%%%%%%%%%%%%%%%%%%%%%%%%%%%%%%%%%%%%%%%%%%%%%%%%%%%%%
%%%%%%%%%%%%%%%%%%%%%%%%%%%%%%%%%%%%%%%%%%%%%%%%%%%%%%%%%%%%%%%%%%%%%%%%%%%%%%%%%%%%%%%%%%%%%%%%%%%%%%%%
\begin{table*}
\begin{center}
\begin{threeparttable}
\caption{Adopted parameters for both the HD\,189733 and 51\,Peg systems.}
\begin{tabular}{ |l|c|c| } 
\hline
Parameter & Value & Reference \\
\hline
\textbf{HD\,189733} \\
$T_\mathrm{eff}$\,(K) & $4875\pm43$ & \citet{Boyajian2015} \\ 
log$(g)$ & $4.56\pm 0.03$ & \citet{Boyajian2015} \\ 
Fe/H & $-0.03\pm0.04$ & \citet{Bouchy2005} \\ 
$K_*\,(\mathrm{m\,s^{-1}})$ & $205\pm6$ & \citet{Bouchy2005} \\ 
$R_*\,(R_\odot)$ & $0.766^{+0.007}_{-0.013}$ & \citet{Triaud2009} \\
$v_\mathrm{sys}$\,(\kms{}) & $-2.361\pm0.003$ & \citet{Bouchy2005} \\
\\
\textbf{HD\,189733\,b} \\
$R_\mathrm{p}\,(R_J)$ & $1.178^{+0.016}_{-0.023}$ & \citet{Triaud2009} \\
\kp\,(\kms{}) & $154^{+14}_{-10}$ & \citetalias{Birkby2013} \\
\\

\textbf{51\,Peg} \\
$T_\mathrm{eff}$ (K) & $5793\pm70$ & \citet{Fuhrmann1997}\\ 
$K_*\,(\mathrm{m\,s^{-1}})$ & $55.65\pm0.53$ & \citet{Wang2002} \\ 
%$R_*\,(R_\odot)$ & $1.237\pm0.047$ & \citet{VanBelle2009} \\
$R_*\,(R_\odot)$ & $1.1609589^{+0.0222188}_{-0.0807567}$ & \citet{GaiaDR22018} \\
$v_\mathrm{sys}$\,(\kms{}) & $-33.2\pm1.5$ & \citetalias{Brogi2013} \\
\\
\textbf{51\,Peg\,b} \\
%$R_\mathrm{p}\,(R_J)$ & $1.9\pm0.3$ & \citet{Martins2015} \\
$R_\mathrm{p}\,(R_J)$ & $1.2$ & See section~\ref{sec:3 gp 51peg} \\
\kp\,(\kms{}) & $134.1\pm1.8$ & \citetalias{Brogi2013} \\

\hline
\label{table:system params}
\end{tabular}
\end{threeparttable}
\end{center}
\end{table*}

\begin{figure*}
    \includegraphics[width=\textwidth]{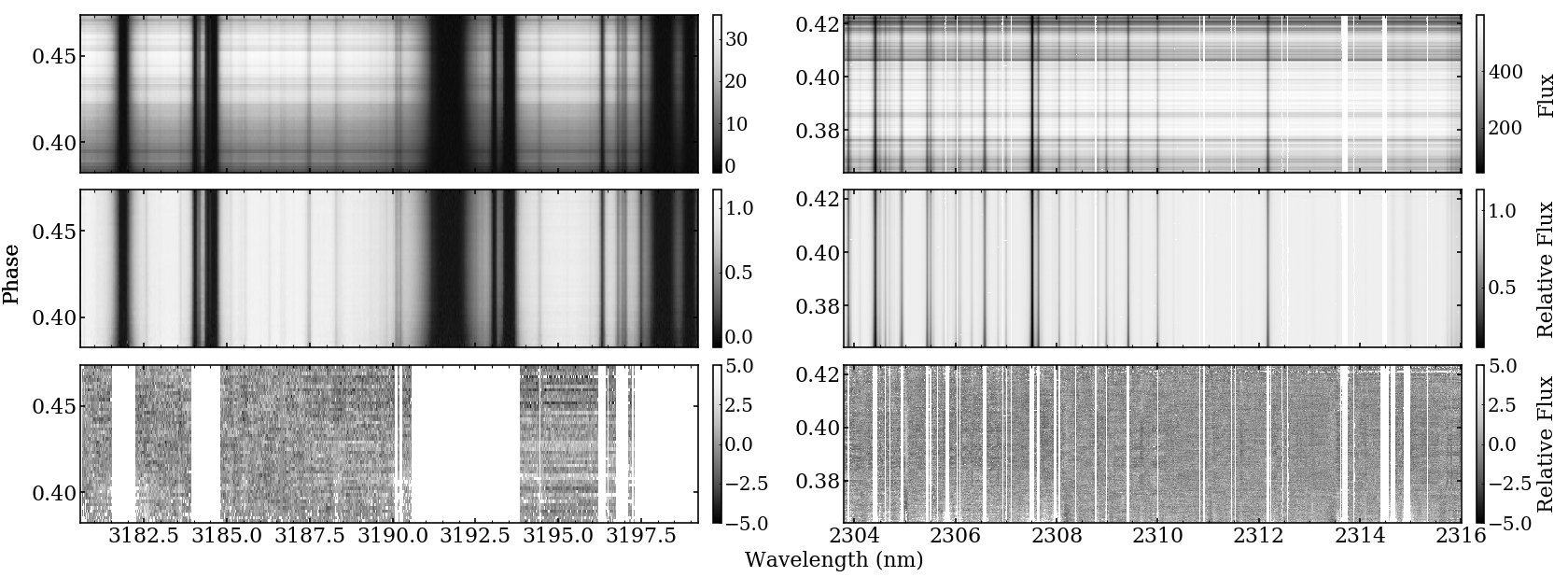}
    \caption{HD\,189733 (\emph{left}) and 51\,Peg (\emph{right}) spectra imaged on CRIRES detectors 1 and 2 respectively. \emph{Top panel:} Reduced spectra matrix from \citetalias{Birkby2013} and \citet{Chiavassa2019} post alignment, background subtraction and bad pixel correction. \emph{Middle panel:} Continuum-normalized spectra, showing vertical bands of telluric absorption. In the case of HD\,189733, the stellar spectrum is removed between this and the next panel. \emph{Bottom panel:} Residuals having removed the GP telluric model and masked appropriate columns, weighted by propagated uncertainties.}
    \label{fig:data-reduction}
\end{figure*}

\subsection{Attempted recovery of the planet signal}
\label{sec:3 recovery of planet spectrum}
\subsubsection{HD\,189733\,b}
\label{sec:3 gp hd189}
HD\,189733 is a K1V-type star with $T_\mathrm{eff} = (4875 \pm 43)\,\mathrm{K}$, therefore we expect few strong stellar lines in the $3.2\,\mu\mathrm{m}$ domain. However, we assessed the telluric-corrected residuals at the locations of expected stellar lines with greater than 5\% absorption and found up to 2.5 times higher dispersion than the rest of the spectrum. In the original study the authors did not explicitly remove any stellar features, as the PCA approach used to remove the tellurics was also able to capture the comparably weaker stellar lines, which shift by less than a pixel during the observations. Since our GP telluric removal assumes all stellar features are removed in order to scale by airmass, we remove them beforehand; we adopt a 1D stellar model from the PHOENIX stellar atmosphere model grid \citep{Husser2013}, interpolating\footnote{All interpolations made using the \textsc{scipy.interpolate} package} between the models for the literature values $T_\mathrm{eff}=4875\,\mathrm{K}$, $\log g = 4.56$ and metallicity [Fe/H]~=~-0.03 (listed in table~\ref{table:system params}). We convolve the stellar model to the CRIRES resolution with a Gaussian kernel, scale according to the depth of two known stellar lines in the observations, and divide it out. Next, synonymous with typical high-resolution cross-correlation analyses, we correct for variation in throughput via continuum normalization; for each spectrum we fit a low-order polynomial to the top 10\% of fluxes in each detector, and divide it out. The normalization is not perfect here due to significant absorption in these spectra. Once normalized, we endeavour to remove the telluric contamination; water bands around $3.2\,\mu\mathrm{m}$ contribute to significant and even saturated telluric absorption, seen as vertical bands in the central, left-hand panel of Fig.~\ref{fig:data-reduction}. We divide out the predictive GP telluric model $T(\lambda, t)$ producing residuals as shown in the bottom, left-hand panel of Fig.~\ref{fig:data-reduction}. Subsequently, we flag any columns of the matrix in which the scatter of the data is systematically larger than the formal uncertainty from the CRIRES pipeline, and choose to inflate the propagated uncertainties accordingly. In some regions the residuals showed a significantly larger scatter than the average, corresponding to regions of saturated tellurics. These regions are masked post telluric removal; we mask a column if its standard deviation is larger than a few factors of the median across the spectrum, with additional sigma clipping of individual points. We were left with an average of 52\% unmasked pixels across detectors 1, 3 and 4 but only 27\% for detector 2 given the severe atmospheric absorption between $3204.45$ and $3222.45\,\mathrm{nm}$. Having applied the masks, the average residual dispersion in continuum regions was $1.4$ times the photon noise level.

We now attempt to recover the previously published planetary atmospheric detection, namely the H$_2$O planetary spectrum detected at $4.8\sigma$ in \citetalias{Birkby2013}. Before training and conditioning the GP for the planetary spectrum on the telluric-corrected residuals, we shift them into the planetary rest frame via equation~(\ref{eqn: vp}) for an assumed $v_\mathrm{sys}~=~-2.361$\,\kms{}, and $v_\mathrm{bary}$ calculated using \textsc{barycorrpy} \citep{Kanodia2018,Wright2014}. We evaluate the GP predictive distribution over a test grid corresponding to one data point per resolution element of the spectrograph. In typical cross-correlation analyses, a detection is often quantified by dividing the values in the summed 2D cross-correlation function by the standard deviation across the entire matrix (or sometimes on a per spectrum basis), thereby providing a signal-to-noise detection significance. Since we do not utilise cross-correlation, we base our definition of a `detection' on others factors: primarily we only consider a detection if the recovered orbital velocity is in agreement with the published (or later injected) velocity. Applying this framework to the observed fluxes, the GP could not detect the true planet signal. We explore two potential reasons why the GP fails to recover the true signal: imperfect telluric removal, impact of which is discussed in section~\ref{sec:3 other detrending}, and the insensitivity of the GP to the noise levels in these data.

\begin{figure}
    \includegraphics[width=0.47\textwidth]{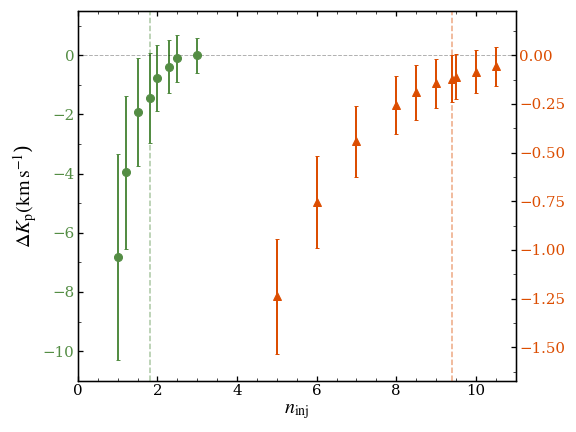}
    \caption{Difference between the injected and recovered \kp{} for HD\,189733\,b (green circles and left axis, injected \kp~$=154$\,\kms{}) and 51\,Peg\,b (orange triangles and right axis, injected \kp~$=134$\,\kms{}). The coloured dashed lines indicate the smallest $n_\mathrm{inj}$ for which we obtain a detection ($n_\mathrm{inj}=1.8$ and $9.4$).}
    \label{fig:Kp-ninj}
\end{figure}

\begin{figure*}
    \includegraphics[width=\textwidth]{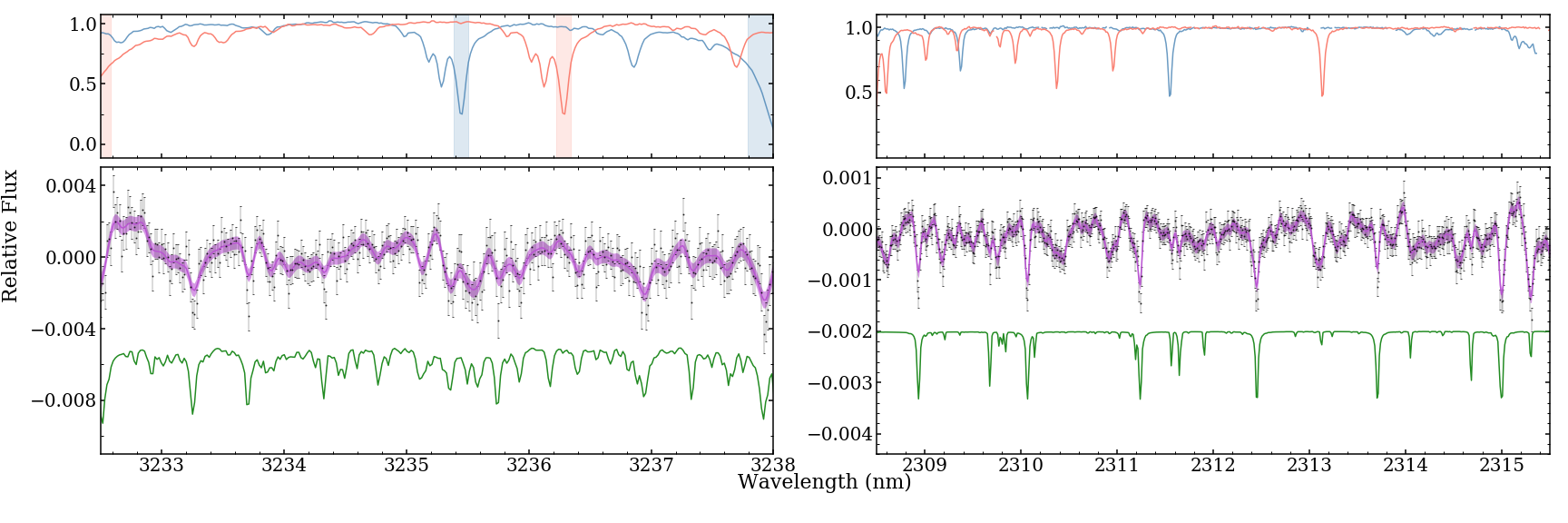}
    \caption{Smallest model injection GP recoveries. \emph{Top panels:} Normalized average spectra plotted in the planet rest frame, dominated by telluric absorption features, with those corresponding to minimum and maximum $v_\mathrm{p}$ (see equation~\ref{eqn: vp}) shown in blue and red respectively. Any masked wavelength channels are indicated by shaded regions. \emph{Bottom panels:} Sub-section of the GP telluric-corrected residual fluxes, binned to CRIRES resolution (again in the planet rest frame) in black, with GP predictive mean and posterior uncertainty ($1\sigma$ interval) overplotted in purple. The injected model spectrum is plotted in green, vertically offset for reference. \emph{Left-hand panels:} Recovered HD\,189733\,b H$_2$O spectrum with injection strength $n_\mathrm{inj}=1.8$. \emph{Right-hand panels:} Result for 51\,Peg\,b, having injected a CO + H$_2$O model with strength $n_\mathrm{inj}=9.4$.}
    \label{fig:smallest-recovery}
\end{figure*}

In applying GP regression we make no assumptions \emph{a priori} with regard to the shape or form of the planetary spectrum, including the locations in wavelength of the absorption lines. The only assumptions we make are regarding the functional form of the covariance between data points, as given in equation~(\ref{eqn: matern32 kernel}). This potentially reduces the sensitivity of the GP method compared to HRCCS, wherein comparisons are made with a pre-defined template of absorption lines. To investigate the strength of signal which could be recovered, we inject model planet spectra at the previously detected radial velocity (\kp{}, $v_\mathrm{sys}$), prior to the data reduction. Cross-correlating an H$_2$O + CO$_2$ model produced the highest detection significance in \citetalias{Birkby2013}, a planet spectrum model produced via line-by-line calculations with reference to the HITEMP line database \citep{Dekok2013,Rothman2010}. We proceed with the H$_2$O-only model, as in \citetalias{Birkby2013}, since the inclusion of CO$_2$ only increased the detection significance by $0.3\sigma$. This model assumes a baseline continuum temperature $T_1=500$\,K with corresponding pressure $P_1=10^{-1.5}$, constant lapse rate and upper atmospheric temperature $T_2=1350$\,K, and VMR(H$_2$O)\,$=10^{-5}$. We scale the model, $F_\mathrm{model}$, according to the host stellar continuum and inject into the observed (pre-processed) spectra, $F_\mathrm{obs}$. We adapt slightly the approach of \citet{Brogi2013, Brogi2014}, and choose to scale the model according to the continuum of the observations, $F_\mathrm{cont}$, to avoid injecting additional telluric (and stellar) features. Our simulated observations are then given by
\begin{equation}
    F_{\mathrm{sim}} = F_{\mathrm{obs}} + F_\mathrm{cont} \Big( n_\mathrm{inj} \frac{F_\mathrm{model}}{F_*}\Big(\frac{R_\mathrm{p}}{R_*}\Big)^2 \Big),
\label{eqn:model-inj}
\end{equation}
where $F_*$ is taken to be the blackbody estimate of the stellar continuum for the stellar effective temperature $T_\mathrm{eff}$. Subsequently, these new spectra are analysed identically to the real spectra. We repeat the GP routine for different $n_\mathrm{inj}$, monitoring the recovered \kp{} and uncertainty (the $1\sigma$ interval of the marginalized posterior distribution), as shown in Fig.\,\ref{fig:Kp-ninj}. We recover a detection at $n_\mathrm{inj}=1.8$. The GP planet spectrum predictive mean and uncertainty for this injection strength is shown in the bottom, left-hand panel of Fig.\,\ref{fig:smallest-recovery} for an unmasked portion of the spectrum, having adopted the MLE values \kp~$=152.55 \pm 1.53 $\,\kms{}, consistent with the injected \kp~$=154$\,\kms{}, and the GP hyperparameters as shown in Fig.\,\ref{fig:mcmc-posteriors}. Scaling the planet model according to equation~(\ref{eqn:model-inj}) with \ninj{}$=1$, the expected line strength (of the deepest lines) is $2.43 \times 10^{-3}$, which corresponds to about 20.1\% of the average standard deviation of the residuals. Note that \citetalias{Birkby2013} used a scale factor $n_\mathrm{inj} = -0.56$ to cancel out the real signal and recover a $0\sigma$ detection, thus computing an H$_2$O line contrast ratio of $(1.3 \pm0.2)\times 10^{-3}$. This considered, $n_\mathrm{inj}=1.8$ corresponds to $\sim 3.2$ times the detected cross-correlation signal strength.

%%%%%%%%%%%%%%%%%%%%%%%%%%%%%%%%%%%%%%%%%%%%%%%%%%%%%%%%%%%%%%%%%%%%%%%%%%%%%%%%%%%%%%%%%%%%%%%%%%%%%%%%
%%%%%%%%%%%%%%%%%%%%%%%%%%%%%%%%%%%%%%%%%%%%%%%%%%%%%%%%%%%%%%%%%%%%%%%%%%%%%%%%%%%%%%%%%%%%%%%%%%%%%%%%

\subsubsection{51\,Pegasi\,b}
\label{sec:3 gp 51peg}
For the solar-type 51\,Peg there are conspicuous, strong stellar CO lines in the vicinity of 2.3\,$\mu$m; the absorption features in the composite spectrum are a convolution of telluric and stellar lines. These were challenging to remove, preventing detection of CO in the original analysis \citepalias{Brogi2013} using spectra from the third observing night (25th October). For this night the planet traversed superior conjunction, resulting in the CO lines of the planet and star overlapping. In the \citet{Chiavassa2019} reanalysis the authors used a more complete 3D stellar model, accounting for stellar convection, rather than the scaled solar 1D stellar model used previously. They then observed an increase in the detection significance of CO when including the spectra from 25th October. In this work we analyse the stellar-corrected residuals of \citet{Chiavassa2019}.

We normalize each spectrum by fitting a low-order polynomial to the top 10\% of pixels within each 10th of the spectrum to simultaneously capture any slight curvature across the order. At $2.3\,\mu\mathrm{m}$ we expect less severe telluric absorption than at $3.2\,\mu\mathrm{m}$, with features predominantly due to CH$_4$, which are known to be less variable than H$_2$O. Having divided out the continuum, and with stellar lines having been previously removed, only telluric absorption features common in wavelength and the Doppler shifting, relative planet atmospheric absorption remain. We remove the predictive GP telluric model, propagating the photon noise uncertainty estimates. The strong telluric line at $\lambda \simeq 2328.5$\,nm, likely a feature of a third species, proved difficult to detrend so we choose to mask it. Additionally, post telluric removal we apply traditional channel masking using the same approach as detailed in Section~\ref{sec:3 gp hd189}.

Combining all three nights of data, once again we do not recover a detection of the true planetary signal. The template planet spectrum we inject is a CO + H$_2$O model based on a non-inverted temperature-pressure profile, with which cross-correlation produced a $5.9\sigma$ detection in \citetalias{Brogi2013}. This model was produced using the HITEMP data base \citep{Rothman2010}, assuming continuum temperature $T_1=1250\,\mathrm{K}$ and pressure $P_1=0.1$\,bar with constant lapse rate to top boundary pressure $P_2 = 1\times 10^{-4}$\,bar, $T_2 = 500$\,K. The volume mixing ratios were VMR(H$_2$O)\,$=3\times 10^{-4}$ and VMR(CO)\,$=1\times 10^{-4}$; it should be noted that these absolute abundances were only weakly constrained since the models were not scaled for the cross-correlation in \citetalias{Brogi2013}. Though the authors also investigated the presence of CH$_4$, they did not observe a significant detection with pure CH$_4$ models. That said, at the time the line lists for CH$_4$ were incomplete and inaccurate for the purposes of high-resolution studies, and it would be important to re-evaluate conclusions regarding the presence of CH$_4$ using updated data bases (e.g. \citealt{Hargreaves2020}).

Scaling the model using equation~(\ref{eqn:model-inj}) we obtain an expected absorption line strength of $1.48\times 10^{-4}$ relative to the continuum level, though again we note this may not be the true line strength of the signal in this data set. A contrast ratio of this scale is much smaller than the residual scatter of the GP telluric-corrected fluxes, corresponding to $3.33\%$ for detector 1. We scale the model spectrum using different values of \ninj{}, inject and process through the GP framework. For the radius of 51\,Peg\,b we adopt the average $R_\mathrm{p}$ of similar mass planets in the NASA exoplanet archive\footnote{https://exoplanetarchive.ipac.caltech.edu/} ($R_\mathrm{p}\sim 1.2\,R_\mathrm{J}$), which agrees with the limits proposed by both \citet{Scandariato2021} and \citet{Birkby2017}. For $n_\mathrm{inj}=9.4$ we recover \kp~$=133.88\pm0.12$\,\kms{}; the right-hand panel of Fig.~\ref{fig:smallest-recovery} shows the resulting GP from a subregion of detector 2 containing several planetary absorption lines, plotted over the telluric-corrected residual fluxes binned to one data point per resolution element. In this formalism $R_\mathrm{p}$ and $R_\mathrm{*}$ act as scaling factors in addition to \ninj{}, and use of a larger planetary radius would translate to a lower \ninj{}. Thus, to clarify, the injected model has an average line strength of $\sim 1.39 \times 10^{-3}$, a similar strength to the real \hdb{} signal. We note that the real signal exists in the data, with a cross-correlation detection uncertainty of $\sim 2$\,\kms{}. Therefore, though we know precisely the velocity at which we inject the model, it may be that the real signal is offset. At low \ninj{} then, it is plausible that the injected and (slightly offset) real signals blend together, distorting the shape of the recovered signal as well as the recovered \kp, therefore impeding a detection.

\subsubsection{Radial velocity semi-amplitude estimation}
\label{sec:3 rv estimate}
As explained in Section~\ref{sec:2 planet fit} we obtain estimates of \kp{} and 2 hyperparameters with this method; Fig.~\ref{fig:mcmc-posteriors} shows that the \kp{} estimates are tightly constrained. Inherently, the GP for the planet component assumes that the data have been perfectly detrended, so what remains is only white noise and the planet signal. This considered, it follows that the theoretical constraining limit of \kp{} is a combination of the spectrograph resolution, SNR, range of phase coverage, and number of (detectable) lines. In reality, however, residual stellar and telluric features exist (with negligible intra-night radial velocity) and become dominant for low \ninj{}, wherein the recovered \kp{} is seen to diverge from the injection velocity (Fig.~\ref{fig:Kp-ninj}). Though it should be noted that the formal uncertainties from HRCCS are likely not the smallest achievable, since the intermediate interpolation steps have an associated noise contribution, our uncertainty estimates for \kp{} are smaller than expected. To proceed to reliable hypothesis testing, en route to enabling standalone GP detections, we would require a more realistic noise model.

We also see a difference in constraining power of \kp{} between the two data sets. From table~\ref{table:system params} we see that the \kp{} estimates from HRCCS analyses have maximal uncertainties of $14$\,\kms{} and $1.8$\,\kms{}, with those for \hdb{} a factor of $\sim 7.8\times$ larger. There are a few conceivable causes for this difference: dynamics and other effects may cause broadening of the spectral lines and thus the radial velocity signature, but the higher signal-to-noise of the \pegb{} data due to wider phase coverage is likely to be a dominant factor. Since these are intrinsic to the data, we therefore expect a similar factor in our analyses. There may have also been slight differences in the methods used to estimate \kp{} from the cross-correlation maps by the authors of \citetalias{Birkby2013} and \citetalias{Brogi2013}. Additionally, the CO lines of \pegb's spectrum are rather more distinct and well defined than the spectrum of \hdb{}, which is likely to significantly aid the detection.

%%%%%%%%%%%%%%%%%%%%%%%%%%%%%%%%%%%%%%%%%%%%%%%%%%%%%%%%%%%%%%%%%%%%%%%%%%%%%%%%%%%%%%%%%%%%%%%%%%%%%%%%
%%%%%%%%%%%%%%%%%%%%%%%%%%%%%%%%%%%%%%%%%%%%%%%%%%%%%%%%%%%%%%%%%%%%%%%%%%%%%%%%%%%%%%%%%%%%%%%%%%%%%%%%

\subsection{Using other detrending algorithms}
\label{sec:3 other detrending}
As mentioned in Section~\ref{sec:3 gp hd189}, poor telluric modelling using GP regression may contribute to inability to detect the buried planet signals. In order to keep the GP model sufficiently simple for the sequential method implemented here, we have only allowed temporal telluric variation corresponding to airmass. In reality telluric behaviour is more complex than this, hence notoriously difficult to model accurately in the NIR, particularly in the case of saturated and non-water-based absorption lines. An additional cause may be the interference of lines from the different components in equation~(\ref{eqn:composite spectrum}). In the case where stellar, telluric and/or planet lines overlap, successive removal will partially affect the other components in the relevant exposures; this issue could potentially be improved with the hypothesized simultaneous GP approach mentioned in Section~\ref{sec:2 full method hrs}. 

To test the contribution of imperfect telluric removal to insensitivity of the GP method to the real signal, we employ the original detrending methods, SYSREM and an airmass detrending approach, which enabled detections via the cross-correlation method. We replicate the detrending methods of \citetalias{Birkby2013} and \citetalias{Brogi2013}, with some slight alterations as detailed in the following sections. Doing so allows us to directly compare the sensitivity of the GP to the cross-correlation approach used in those publications for identifying the planet signal, by eliminating the impact of the GP telluric removal.

%%%%%%%%%%%%%%%%%%%%%%%%%%%%%%%%%%%%%%%%%%%%%%%%%%%%%%%%%%%%%%%%%%%%%%%%%%%%%%%%%%%%%%%%%%%%%%%%%%%%%%%%

\subsubsection{SYSREM}
\label{sec:3 sysrem} % here only apply sysrem to hd189
Compared to the rapidly Doppler shifting planetary signal (e.g. 103-24\,\kms{} for the HD\,189733\,b observations) the telluric and stellar spectra appear quasi-stationary, experiencing only sub-pixel shifts over the course of an observing night. Analogous to other SYSREM implementations, we divide through the mean flux in each pixel and subtract unity, leaving only residual temporal variation \citepalias{Birkby2013}. We note that it is necessary to apply masks to regions of saturated telluric absorption prior to input; for these tests we mask the same columns of the spectral matrix as \citetalias{Birkby2013}. We run the SYSREM algorithm for the same number of iterations, $N_\mathrm{SYSREM}$, for detector 1 and 3 as used in \citetalias{Birkby2013} ($N_\mathrm{SYSREM}=8$ and 1 iteration respectively). We also include detectors 2 and 4, unused in \citetalias{Birkby2013}, employing $N_\mathrm{SYSREM}=5$ for each. We observe a non-detection for the SYSREM-treated fluxes. Separately, we test $N_\mathrm{SYSREM}$ found using our own optimization process (explained further in Section~\ref{sec:4 sysrem tuning}), but still do not achieve a detection. We inject the H$_2$O model spectrum in the same way as Section~\ref{sec:3 gp hd189} and again investigate the lowest $n_\mathrm{inj}$ for which we observe a detection. At $n_\mathrm{inj}=1.4$ we recover an agreeable planet orbital velocity, \kp~$=155.62\pm1.76$\,\kms{}. SYSREM visibly better suppresses the noise, thus it is unsurprising that smaller signals are detectable.

%%%%%%%%%%%%%%%%%%%%%%%%%%%%%%%%%%%%%%%%%%%%%%%%%%%%%%%%%%%%%%%%%%%%%%%%%%%%%%%%%%%%%%%%%%%%%%%%%%%%%%%%
%%%%%%%%%%%%%%%%%%%%%%%%%%%%%%%%%%%%%%%%%%%%%%%%%%%%%%%%%%%%%%%%%%%%%%%%%%%%%%%%%%%%%%%%%%%%%%%%%%%%%%%%
\subsubsection{Airmass-detrending approach}
\label{sec:3 airmass-detrending}
\begin{figure}
    \includegraphics[width=0.5\textwidth]{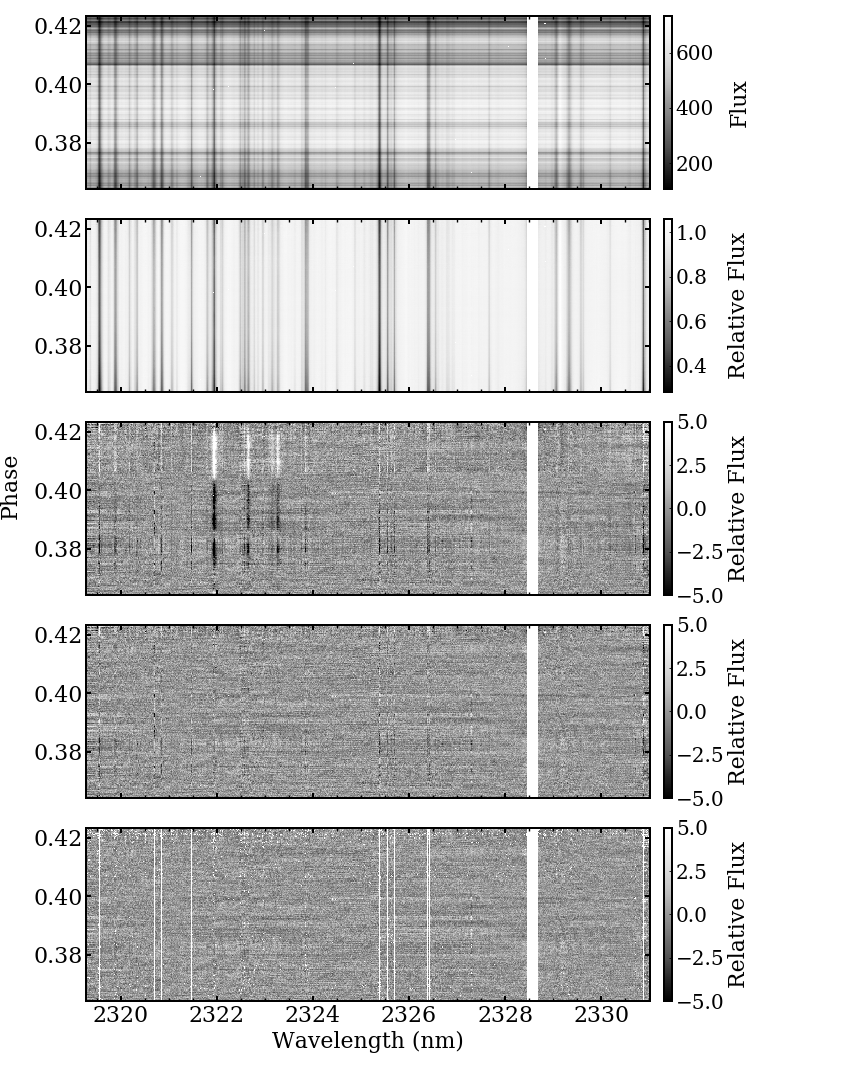}
    \caption{Data reduction of 51\,Peg spectra from October 16th imaged on CRIRES detector 3. The strong telluric line at $\lambda \simeq 2328.5$\,nm is masked throughout. \emph{Top panel:} Stellar-subtracted fluxes from \citet{Chiavassa2019}. \emph{Second panel:} Continuum-normalized fluxes. The latter three panels show the results of successive routines of the linear regression telluric removal technique outlined in Section~\ref{sec:3 airmass-detrending}. Noisy channels are masked post detrending in the final panel.}
    \label{fig:51peg-spectra}
\end{figure}

We employ the linear regression technique presented and used by \citet{Brogi2012,Brogi2014} and \citet{Schwarz2015} to remove the telluric lines imprinted on the 51\,Peg spectra. After normalizing the spectra, we identify and remove any airmass-related variation in each pixel before sampling the residuals at user-defined $\lambda_s$, known strong telluric line positions, for any second order effects, and applying a high-pass filter as discussed in Section~\ref{sec:1 intro}. These steps replicate the approach outlined in \citetalias{Brogi2013}, with the exception of normalizing by column variances, a final step commonly used in cross-correlation analyses to `down-weigh' particularly noisy pixels. We choose to omit this step since it alters the variance of the data. Instead, we mask noisy channels and apply sigma clipping during the final high-pass filtering using the masking algorithm described in Section~\ref{sec:3 gp hd189}. Fig.~\ref{fig:51peg-spectra} shows the spectral matrix residuals from a single night (October 16th) after each step of the reduction, with the third panel showing the residuals having first removed the airmass trend. Again, we do not achieve a detection of the planet signal having treated the fluxes with this sequence of detrending algorithms. We varied $\lambda_s$ from those used in \citetalias{Brogi2013} but to no avail. Having injected the CO + H$_2$O model spectrum with $n_\mathrm{inj}=8.5$, we achieve a marginal posterior mean \kp~$=134.13\pm 0.14$\,\kms{}.

The ability of the GP to recover a marginally weaker planet signal both here and using SYSREM suggests that the GP telluric model is underperforming, though we note that here the aforementioned $n_\mathrm{inj}=8.5$ corresponds to an average SNR$_\mathrm{line}\sim0.36$ (of the strongest lines) when compared to the noise of the detrended residuals, higher than SNR$_\mathrm{line}\sim0.29$ for the GP-treated fluxes. Significant residuals are visible in the channels containing stronger telluric lines after the first pass of this linear regression detrending technique (third panel of Fig.~\ref{fig:51peg-spectra}), illustrating that the behaviour of tellurics is more complex than assumed. This further supports that the airmass scaling assumption is likely to be too simplistic and the predominant shortfall of the GP sequential method for detecting small signals.

\section{Quantitative Analysis of the Impact of Detrending}
\label{sec:4 impact of telluric methods}
It has been shown that detrending methods used in past HRCCS analyses can alter the intrinsic planet signal. So, in order to avoid biasing the exoplanet atmospheric retrieval, it is important to replicate any alterations by subjecting the cross-correlation template to the same detrending process \citep{Brogi2019,Serindag2021}. Still, it is difficult to identify the exact effect of detrending as the planet spectrum is unseen. Alternatively, masks can be applied prior to detrending, to protect regions expected to contain planet absorption or emission features \citep{Schwarz2015}. Not only does this require input knowledge of atmospheric content, but it is also non-ideal to rid parts of the domain which may contain valuable information. Our methodology allows us to observe and measure the modifications to the planet spectrum itself.

\subsection{Defining detection metrics}
\label{sec:4 detection metrics}
We develop four metrics to assess the impact of the telluric removal on the planet signal, having considered which properties of a spectrum are important to preserve for the purposes of retrievals. When characterising a planetary atmosphere, we need to distinguish features in absorption from those in emission, indicative of the presence of temperature inversions in the upper atmosphere. Since here we are working with injected spectra which are known to contain only absorption features we reserve this assessment for future work. When comparing the recovered `planetary spectrum' to the injected model spectrum, we first assess if absorption features are observed at the same wavelength locations. Then, we measure the depths of absorption lines relative to the `pseudo-continuum', which remains after detrending. These are important for deciphering atomic and molecular abundances, and are also affected by sources of broadening such as rotation. 
We identify any features greater than 2$\sigma$ away from the continuum in the injected model spectrum, and fit the located absorption lines with a Gaussian profile to measure their depths, widths and positions. We note that dynamics in the planetary atmosphere can distort spectral features, resulting in asymmetric profiles. For the test cases in this work we observe little distortion of the line profiles in the injected models, therefore, while we measure the full width at half maximum (FWHM) of the lines, we leave assessment of line shape to future work and assume symmetric features. In addition, we are interested in reproducing the planet `pseudo-continuum', thus minimizing the presence of spurious features in the telluric-corrected fluxes. Therefore, as a third metric we select regions of continuum and assess dispersion. We compute a final combined metric which considers that each metric should be as close to unity as possible, weighting by uncertainty where appropriate and normalizing before combining. To summarize, we measure the following four properties: 
\begin{enumerate}
    \item line depth;
    \item line FWHM; 
    \item dispersion in continuum regions;
    \item a metric that combines the above.
\end{enumerate}
For the model comparison, we inject the template spectrum into a white noise matrix with a dispersion set to the photon noise level, at the published radial velocity of the planet. Then, the model comparison spectral properties are those measured having shifted the `noise + injected model' matrix to the planet rest frame and binned to the resolution of CRIRES.

%%%%%%%%%%%%%%%%%%%%%%%%%%%%%%%%%%%%%%%%%%%%%%%%%%%%%%%%%%%%%%%%%%%%%%%%%%%%%%%%%%%%%%%%%%%%%%%%%%%%%%%%
%%%%%%%%%%%%%%%%%%%%%%%%%%%%%%%%%%%%%%%%%%%%%%%%%%%%%%%%%%%%%%%%%%%%%%%%%%%%%%%%%%%%%%%%%%%%%%%%%%%%%%%%

\subsection{Tuning detrending parameters}
\label{sec:4 sysrem tuning}
Though SYSREM is a powerful detrending tool, it can be easy to degrade the planetary signal. In most cases it is likely that planet spectral features experience sub-pixel shifts between exposures - SYSREM can begin to remove these sub-pixel common modes once the telluric spectrum has been removed for higher (user-specified) number of iterations. There is therefore a fine balance between removing as much of the stellar and telluric contamination as possible while retaining the planet signal. The common method used in the literature to determine the optimal number of SYSREM iterations, $N_\mathrm{SYSREM}$, is to inject a model planet spectrum at the expected radial velocity (\kp{}, $v_{\mathrm{sys}}$), and select $N_\mathrm{SYSREM}$ that corresponds to the maximum recovery significance, typically the SNR of the peak in a cross-correlation map \citep{Birkby2017,Nugroho2017,Kesseli2020}. The noise is estimated as the standard deviation of CCF values in a user-defined region, located away from the predicted peak. There have been some concerns that this SNR is not a robust metric for the purpose of tuning detrending parameters; \citet{Cabot2019} showed use of SNR as an optimization metric to be susceptible to false-positive detections due to the presence of spurious features in the cross-correlation maps. This has also created caution over optimization of iteration number on an order-by-order basis, leading some to opt for uniform SYSREM use across spectral orders \citep{Nugroho2017,Nugroho2020}. 

\begin{figure}
    \includegraphics[width=0.45\textwidth]{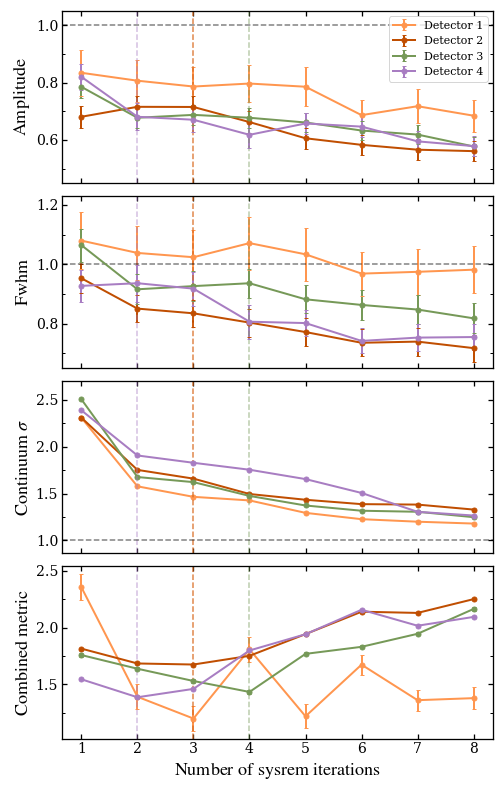}
    \caption{Assessment of planet (CO + H$_2$O) spectrum recovery for 51\,Peg\,b observations taken on October 16th using different number of SYSREM iterations, having injected a signal of strength $n_\mathrm{inj}=14.0$. Each of the three metrics, (i)-(iii) in Section~\ref{sec:4 detection metrics}, is shown individually in the first three rows. \emph{Bottom panel:} Minimization of combined metric having considered line amplitude, FWHM, and continuum standard deviation recovery of the planet spectrum, as compared to the injected (noisy) model. Optimization is performed on each detector independently as indicated by the different colours.}
    \label{fig:51peg-optimal-sysrem}
\end{figure}

To determine the optimal number of SYSREM iterations for recovering the planet signal, we first inject a signal larger than the strength detected in Section \ref{sec:3 sysrem}. We train the GP only on the residuals following three iterations applied to each detector. We then run SYSREM for eight iterations, saving the residuals after removal of each principal component and evaluate the GP, having used the globally optimized covariance function and \kp{}, for each residual-flux matrix. We make use of the three metrics developed in Section \ref{sec:4 detection metrics} to assess line recovery of the deepest planet absorption lines. The amplitude and FWHM recovery-injection ratios are averaged over all the selected absorption lines. In the case of the 51\,Peg\,b data set we repeat this process for each night separately; it is expected that telluric contamination will not only vary between detectors but also significantly between nights of observations. The measured properties for the first night of 51\,Peg\,b observations (October 16th) are shown in Fig.~\ref{fig:51peg-optimal-sysrem}. We select the $N_\mathrm{SYSREM}$ that minimizes the combined metric (bottom panel of Fig.~\ref{fig:51peg-optimal-sysrem}). In the case of slight ambiguity concerning the minimum difference between injection and recovery, we select the lower $N_\mathrm{SYSREM}$. The optimum values used going forward are summarised in table~\ref{table:observations}. Since the method presented here does not use the SNR of cross-correlation maps as a detection metric, it bypasses the issue of noise optimization, thus reducing the risk of biasing the detection by optimization of SYSREM for each detector separately.

%%%%%%%%%%%%%%%%%%%%%%%%%%%%%%%%%%%%%%%%%%%%%%%%%%%%%%%%%%%%%%%%%%%%%%%%%%%%%%%%%%%%%%%%%%%%%%%%%%%%%%%%
%%%%%%%%%%%%%%%%%%%%%%%%%%%%%%%%%%%%%%%%%%%%%%%%%%%%%%%%%%%%%%%%%%%%%%%%%%%%%%%%%%%%%%%%%%%%%%%%%%%%%%%%
\begin{figure*}
    \includegraphics[width=\textwidth]{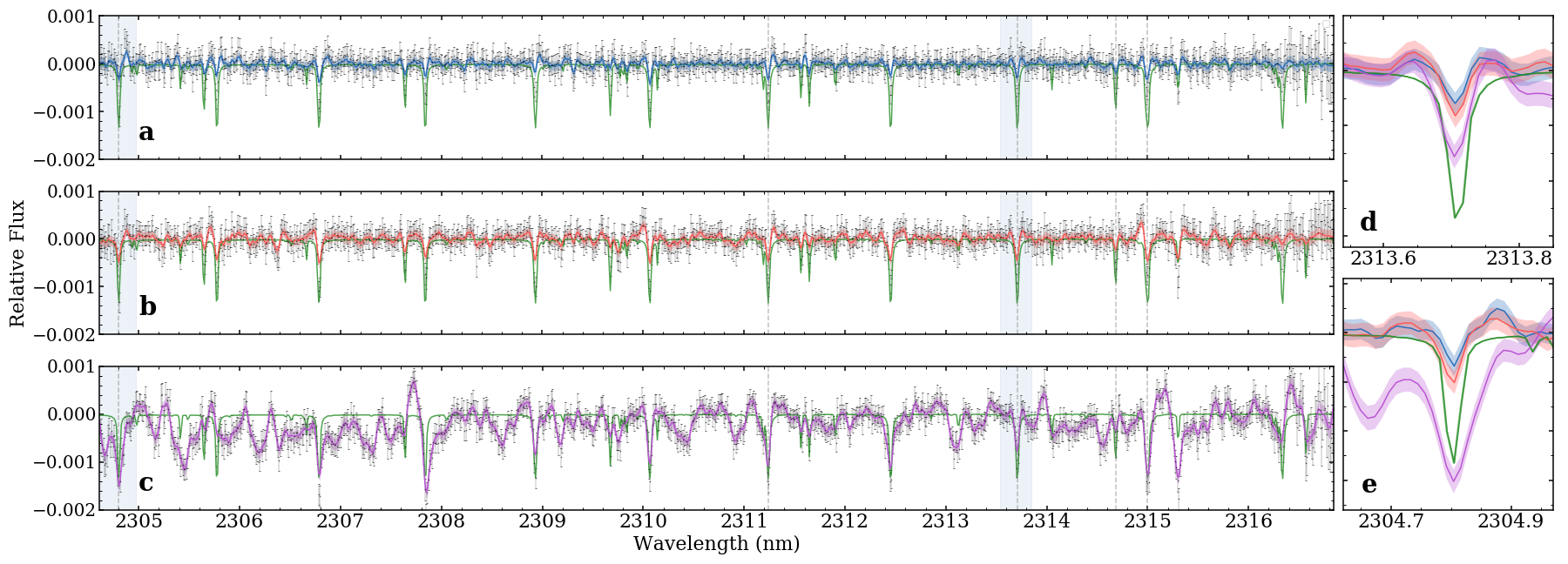}
    \caption{Binned linear regression-corrected 51\,Peg residuals (panel a), SYSREM-corrected fluxes (panel b) and having removed a GP telluric model (panel c), in the planetary rest frame, with injected CO model of strength $n_\mathrm{inj}=9.5$ (shown in green for reference). The GP regressed on the (full set of) residuals is overplotted in each case. The full resolution residuals have a large dispersion, thus to maintain clarity they are not shown here. The dashed lines indicate the locations of the lines given in table~\ref{table:comparing-tellcorr}. Please note that panels (a-c) show three separate rather than sequential processes. Panels (d) and (e) are subsections of the wavelength domain around $\lambda\simeq2313.709$ and $\lambda\simeq2304.804$\,nm, respectively, showing the GP predictive mean regressed on the 3 sets of residual fluxes. The corresponding regions are shaded in blue in panels a-c.}
    \label{fig:51peg-comparing-tellcorr}
\end{figure*}

\subsection{Comparison of tellurics correction methods}
\label{sec:4 comparing telluric removal}
To compare the three methods of telluric removal, we process the same set of fluxes, with a model injection of the same strength, and ensure the masking is identical. \citet{Cabot2019} showed the impact of variation in mask level on detection significance - we do not investigate this explicitly in this section. We run the SYSREM algorithm for an optimal number of iterations as indicated in table~\ref{table:observations} removing each systematic component successively; the optimization process is explained in Section \ref{sec:4 sysrem tuning}. We treat spectra from each detector (and each night) independently. Some authors opt to divide the fluxes through by a global model, the summation of all the identified principal components \citep{Gibson2020,Merritt2020}. We do not observe a significant difference in the recovered planet spectrum using this method instead, at these injection strengths.

Both SYSREM and the airmass fitting approach outperform the GP telluric model in terms of removal of stationary contaminating features, showing lower dispersion in the telluric-corrected residuals, and consequently increasing the ratio of expected line contrast to noise. The HD\,189733 observations suffered saturated telluric absorption for a large proportion of the wavelength domain, up to ~70\% for detector 2. All three methods struggled to model these and we had to apply significant masking. Despite the reduced spectral coverage, there were a few regions containing several absorption features (covering ranges of 1-2\,$\mu$m) that enable comparison of line recovery. As seen in Section~\ref{sec:3 sysrem}, SYSREM enabled recovery of a smaller (injected) planetary signal than the GP sequential method, and similarly using the linear regression method with the \peg{} spectra (Section~\ref{sec:3 airmass-detrending}). That said, we observe a clear reduction in amplitude of the line features in the corrected fluxes when compared to the original injected model amplitudes. To quantitatively compare the impact of the three telluric detrending methods used, we measure the aforementioned line properties after applying the telluric correction, shifting the corrected fluxes to the planet rest frame by the MLE \kp{} (obtained having implemented our GP modelling of the planet spectrum), and binning the residuals to one data point per spectrograph pixel. We show an example result applying each method to the 51\,Peg spectra in Fig.~\ref{fig:51peg-comparing-tellcorr} for $n_\mathrm{inj}=9.5$, which includes the regressed GP overplotted for interest. For this example, the recovered line metrics for some notable lines and values averaged across the 10 deepest lines, are given in table~\ref{table:comparing-tellcorr}, along with the measured continuum standard deviation of a set of pre-defined spectral regions.

Both SYSREM and the airmass linear regression methods perform poorly with regard to maintaining the strength of absorption lines. With the overall variance as a metric, sub-pixel shifts will be identified and removed with these methods; thus, the wings and depths of lines are degraded. Meanwhile, generally we find that the GP telluric model enables better retention of the line depths, and the GP telluric correction reproduces a noisy continuum, greater than three times the standard deviation of the `noise + injected model' for the example given in panel (c) of Fig.~\ref{fig:51peg-comparing-tellcorr} ($n_\mathrm{inj}=9.5$). In addition to averaging these metrics across the spectrum, we assess the wavelength dependence. We identify planet absorption lines that satisfy differing conditions: those that consistently lie in regions with high atmospheric transparency (>90\%) and those that Doppler shift across strong telluric lines (<80\%), traversing the same spectral region in 10-90\% of exposures.

Measuring the amplitudes and FWHM ratios for the lines in each subgroup, we observe that SYSREM and the airmass linear regression residuals cultivate wavelength-independent line recovery, producing consistent ratios in regions of continuum and regions of strong telluric absorption. This is not the case with the recovered lines following the GP telluric correction. Continuing with the example of Fig.~\ref{fig:51peg-comparing-tellcorr} and table~\ref{table:comparing-tellcorr}, of the 10 absorption lines measured, four lines are located in regions with more than 95\% telluric transmission: $\lambda\simeq2311.241$\,nm, $2313.709$\,nm, $2314.683$\,nm, and $2314.999$\,nm. For reference, a portion of the average telluric spectrum for this detector can be seen in the top right panel of Fig.~\ref{fig:smallest-recovery}, plotted in the planetary rest frame to show relative position for the full duration of observations. Shown in the bottom right panel, the line located at $\lambda\simeq2304.804$\,nm traverses a deep telluric line over more than 60\% of the total exposures. The recovered to injected ratios for each of these individual lines are included in table~\ref{table:comparing-tellcorr}. While the example of detector 2 for the 51\,Peg spectra is highlighted here, it is representative of results observed from other detectors. In general, we find that planet spectral lines that consistently lie in regions of light telluric absorption are well recovered, whereas the recovered line profiles are distorted if deep telluric lines are nearby.

We conclude that planet lines that are not in close proximity to telluric features in a large number of exposures are more reliably and accurately recovered with the sequential GP method, compared to the other two detrending methods used. One plausible use of these findings would be to incorporate some GP modelling in an HRS data reduction pipeline, for the purpose of informing the detrending rather than implementing it. For example, one could take the average line depth measured using the GP framework, having excluded lines adjacent to telluric lines, and use it as a scaling factor for the recovered planetary spectrum. This would then combine the strengths of both approaches to extract the most accurate planetary spectrum possible. This is only a proposal here; we stress that our aim in this work is not to put forward a competitive, stand-alone telluric detrending method, but to test a sequential GP framework with the aim of future extension to a simultaneous GP modelling method.

\begin{table*}
\begin{threeparttable}
\caption{Line recovery having implemented different methods of telluric correction. All metrics given as the ratio between those measured in the binned telluric-corrected fluxes and in the injected CO model. Example given for 51\,Peg\,b, detector 2, $n_\mathrm{inj}=9.5$ (all observing nights).}
\begin{tabular}{|l|c|c|c|} 
\hline
Line $\lambda$ (nm) & Airmass linear regression & SYSREM & GP telluric correction \\
\hline
\textbf{Amplitude} \\
Average & $0.629\pm0.038$ & $0.680\pm0.038$ & $1.138\pm0.053$ \\
$2304.804$ & $0.833\pm0.180$ & $0.919\pm0.139$ & $1.575\pm0.227$ \\
$2311.241$ & $0.655\pm0.154$ & $0.692\pm0.118$ & $1.073\pm0.194$ \\
$2313.709$ & $0.576 \pm0.121$ & $0.628\pm0.095$ & $0.923\pm0.125$ \\
$2314.683$ & $0.652\pm0.164$ & $0.636\pm0.184$ & $0.716\pm0.247$ \\
$2314.999$ & $0.534\pm0.108$ & $0.549\pm0.103$ & $1.157\pm0.146$ \\
\\
\textbf{FWHM} \\

Average & $0.739\pm0.049$ & $0.808\pm0.047$ & $1.123\pm0.065$ \\
$2304.804$ & $0.751\pm0.152$ & $0.805\pm0.131$ & $1.185\pm0.239$ \\
$2311.241$ & $0.724\pm0.191$ & $0.758\pm0.147$ & $1.219\pm0.264$ \\
$2313.709$ & $0.603\pm0.133$ & $0.637\pm0.102$ & $0.741\pm0.120$ \\
$2314.683$ & $0.941\pm0.241$ & $0.923\pm0.269$ & $0.801\pm0.355$ \\
$2314.999$ & $0.730\pm0.160$ & $0.795\pm0.147$ & $1.264\pm0.174$ \\
\\
\textbf{Continuum standard deviation} \\

& $1.622$ & $1.726$ & $3.747$ \\
\hline
\label{table:comparing-tellcorr}
\end{tabular}
\end{threeparttable}
\end{table*}

%%%%%%%%%%%%%%%%%%%%%%%%%%%%%%%%%%%%%%%%%%%%%%%%%%%%%%%%%%%%%%%%%%%%%%%%%%%%%%%%%%%%%%%%%%%%%%%%%%%%%%%%
%%%%%%%%%%%%%%%%%%%%%%%%%%%%%%%%%%%%%%%%%%%%%%%%%%%%%%%%%%%%%%%%%%%%%%%%%%%%%%%%%%%%%%%%%%%%%%%%%%%%%%%%
\section{Conclusions}
\label{sec:5 conclusions}
In this work, we have proposed a novel technique for analysis of high-resolution spectra, using Gaussian process regression to sequentially model the telluric and planet spectra directly. The Bayesian approach affords an estimate of planet orbital velocity in addition to the planet spectrum estimate, both with robust uncertainties. We made a number of simplifying assumptions for our models in order to attain efficient computation, notably only allowing airmass variation of the telluric model. Nevertheless, the simplistic, sequential framework has enabled useful tests en route to construction of a comprehensive GP treatment. Though we produced no detection on real data sets, we successfully recovered planet absorption line positions and shapes for injected signals slightly larger than the nominal detected signals from cross-correlation analyses. The method here was developed specifically for dayside exoplanet spectra; however, a similar procedure could be used for transit observations, with alterations considering the Rossiter-Mclaughlin and other stellar effects.

It is clear that the efficacy of the GP routine to retrieve the planet signal is dependent on the performance of the telluric removal for the particular data sets in this work. We investigated the impact of different detrending methods on injected planetary signals, assessing recovered absorption line profiles. Standard techniques, SYSREM and the aforementioned linear regression method, consistently degrade the sought planetary signal, both truncating the absorption features and the wings of lines, which in the most severe case could prevent detection of weaker lines. Any high-resolution analyses should apply these techniques with caution and take care to mimic the effect of detrending when comparing to models. Though SYSREM is very effective at removing the telluric imprint, it has the potential to be quite aggressive, thus requiring careful tuning. It is then difficult to choose the optimum number of iterations to remove as much of the telluric contamination as necessary, to reveal the `true' planet signal while completely retaining that signal. Additionally, we observed that all absorption lines were uniformly affected across the spectrum with both standard detrending methods. In contrast, the line recovery of the combined sequential GP methodology was wavelength dependent, fairing better in regions of greater telluric transparency compared to regions in close proximity to strong telluric lines. The downfall is likely to be the simplistic airmass scaling of the telluric model.

Use of GP regression in this context offers a number of advantages; we obtained detections without the requirement of any prior (exact) knowledge concerning the shape or form of the spectra for construction of a parametrized model. Acquiring an estimate of the planet spectrum is greatly beneficial, and may further help with high-resolution atmospheric retrievals. We note that our current method requires continuum normalization in the data reduction; thus, we do not retain the absolute planetary continuum. This could potentially be rectified via extension to a simultaneous GP fitting. An additional benefit of the method we presented is the natural acquisition of a (GP) likelihood function, thus disposing the need for a conversion tool to output a statistically meaningful detection. Though for the signal-to-noise of these data sets our current method does not have sufficient sensitivity to recover the real planet signals, the prospects of this technique are promising. With the proposed advancements, GPs may offer a data-driven, potential route forward in analysis of high-resolution spectra.

%%%%%%%%%%%%%%%%%%%% ACKNOWLEDGEMENTS %%%%%%%%%%%%%

\section*{Acknowledgements}
Based on observations collected at the European Southern Observatory under ESO programme 186.C-0289. We thank the anonymous referee for their comments towards improving the clarity of this manuscript. A.\,M. acknowledges support from the UK Science and Technology Facilities Council (STFC). S.\,A. and J.\,L.\,B. acknowledge funding from the European Research Council (ERC) under the European Union’s Horizon 2020 research and innovation programme under grant agreement numbers 865624 and 805445. This research has made use of the NASA Exoplanet Archive, which is operated by the California Institute of Technology, under contract with the National Aeronautics and Space Administration under the Exoplanet Exploration Program.
% Entry for the table of contents, for this guide only
%\addcontentsline{toc}{section}{Acknowledgements}

%%%%%%%%%%%%%%%%%%%% DATA AVAILABILITY %%%%%%%%%%%%%
\section*{Data Availability}
The data underlying this article are publicly available through the ESO Science Archive Facility.


\bibliographystyle{mnras}
\bibliography{references} % if your bibtex file is called example.bib
%\renewcommand\bibname{References}

%%%%%%%%%%%%%%%%% APPENDICES %%%%%%%%%%%%%%%%%%%%%
\appendix
\section{Parameter posterior distributions}
\begin{figure*}%
    \centering
    \subfloat{{\includegraphics[width=0.48\textwidth]{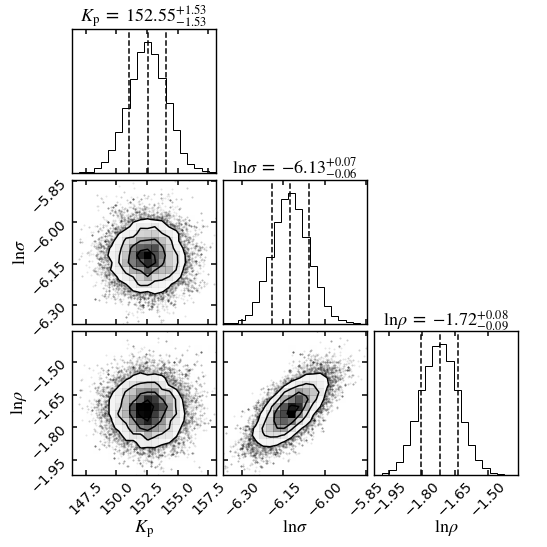} }}%
    \qquad
    \subfloat{{\includegraphics[width=0.48\textwidth]{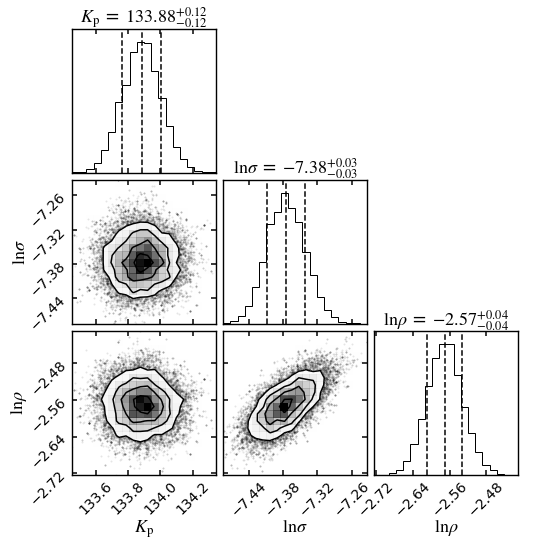} }}%
    \caption{Posterior distributions over radial velocity semi-amplitude \kp{} and GP hyperparameters from the MCMC, for a GP trained on GP telluric-corrected \hd{} (\emph{left-hand panel}) and \peg{} (\emph{right-hand panel}) residuals, with planet spectrum model injections of strengths $n_\mathrm{inj}=1.8$ and $9.4$ respectively. For each panel, the marginalized distributions for each parameter are given on the diagonal, with dashed lines indicating the locations of the 16th, 50th and 84th percentiles.}
    \label{fig:mcmc-posteriors}%
\end{figure*}
%%%%%%%%%%%%%%%%%%%%%%%%%%%%%%%%%%%%%%%%%%%%%%%%%%

% Don't change these lines
\bsp	% typesetting comment
\label{lastpage}
\end{document}